\documentclass[12pt]{article}

\RequirePackage[OT1]{fontenc}
\usepackage{amsmath}
\usepackage{mathabx}
\usepackage{graphicx,psfrag,epsf}
\usepackage{enumerate}
\usepackage{natbib}
\usepackage{url} 
\usepackage{appendix}
\usepackage{ragged2e}

\newcommand{\blind}{1}

\usepackage{amsthm}
\usepackage[hidelinks]{hyperref}
\usepackage{setspace}
\usepackage{pdflscape}
\usepackage{afterpage}
\usepackage[top=1in,right=1.25in,left=1.25in,bottom=1in]{geometry}

\usepackage{url} 
\usepackage{threeparttable}
\usepackage{booktabs}
\usepackage{multirow}
\usepackage{bm}

\newcommand{\specialcell}[2][c]{%
  \begin{tabular}[#1]{@{}c@{}}#2\end{tabular}}

\theoremstyle{plain}

\begin{document}
\RaggedRight
\setlength\parindent{1.5em}

\def\spacingset#1{\renewcommand{\baselinestretch}%
{#1}\small\normalsize} \spacingset{1}


\if1\blind
{
  \title{\bf Resampling Methods with Imputed Data}
  \author{\large Michael W. {Robbins}, Lane Burgette, and Sebastian Bauhoff 
\thanks{Michael W.~Robbins and Lane Burgette are Senior Statisticians with the RAND Corporation, Pittsburgh, PA 15213 (E-mail:~{\em mrobbins@rand.org}, {\em burgette@rand.org}). Sebastian Bauhoff is Assistant Professor of Global Health and Economics (E-mail:~{\em sbauhoff@hsph.harvard.edu}) with Harvard T.~H.~Chan School of Public Health, Boston, MA 02115.  The authors acknowledge funding from grant R21AG058123 from the National Institutes of Health.}
}
  \maketitle
} \fi

\if0\blind
{
  \bigskip
  \bigskip
  \bigskip
  \begin{center}
    {\LARGE\bf Resampling Methods with Imputed Data}
\end{center}
  \medskip
} \fi

\bigskip
\begin{abstract}
Resampling techniques have become increasingly popular for estimation of uncertainty in data collected via surveys. Survey data are also frequently subject to missing data which are often imputed. This note addresses the issue of using resampling methods such as a jackknife or bootstrap in conjunction with imputations that have be sampled stochastically (e.g., in the vein of multiple imputation).  It is illustrated that the imputations must be redrawn within each replicate group of a jackknife or bootstrap. Further, the number of multiply imputed datasets per replicate group must dramatically exceed the number of replicate groups for a jackknife. However, this is not the case in a bootstrap approach. A brief simulation study is provided to support the theory introduced in this note.
\end{abstract}

\noindent%
{\it Keywords:} Missing Data, Multiple Imputation, Jackknife, Bootstrap.

\newtheorem{proposition}{Proposition}
\newtheorem{lemma}{Lemma}
\newtheorem{theorem}{Theorem}
\newtheorem{corollary}{Corollary}
\def\convD{\,{\buildrel D \over\to}\,}
\def\convP{\,{\buildrel p \over\to}\,}
\def\equalD{\,{\buildrel D \over =}\,}

\def\spacingset#1{\renewcommand{\baselinestretch}%
{#1}\small\normalsize} \spacingset{1.5}

\section{Introduction} \label{sec1}

The problem of missing values confounds most efforts to analyze survey data, with a wealth of literature and software developed to address it. Multiple imputation \citep{rubin96, little02}, in which each missing value is independently ``filled in'' finitely many times with plausible draws from a predictive distribution to create several completed datasets has become a preferred manner of accounting for missingness. Although the use of a single imputed dataset in lieu of multiple imputed ones may not inherently bias estimates derived from the data, information can be combined across multiple imputed datasets to adjust estimators of uncertainty for error in the imputations; Rubin \cite{rubin87a} introduced formulas for doing so that are now well-known.

However, any issues presented by missing data are confounded in the anlaysis of surveys since traditional algebraic formulas for variance estimation may not be valid in the presence of a complex survey design. Although linearization \citep{binder83, lumley04} may be used to account for facets like inverse probability weighting and stratification, these formulas have been shown to fail in certain circumstances \citep[e.g.,][]{robbins21}, leading resampling techniques (such as a jackknife or a bootstrap) to be the preferred method of variance estimation for many survey methodologists. Furthermore, additional literature has shown that traditional rules for combining data from multiple imputed datasets fail when used with complex surveys \citep{kott95, kim06}.

Despite the ubiquity of missingness in survey data and the popularity of resampling techniques for analyzing such data, there is limited literature exploring the use of resampling techniques within the context of multiply imputed data. This note attempts to fill in this gap in the literature.

Specifically, we illustrate that missing values should be independently imputed several times within each replication group of a jackknife or bootstrap. This contradicts existing literature which stipulates that each replicate group should be imputed only once \citep[e.g.,][]{little02, miller11}. Furthermore, the number of imputed datasets required for each group increases with the number of replicate groups for the jackknife; however, this is shown to not be the case for the bootstrap. Nonetheless, simulations show that mean squared error is slightly smaller for the jackknife than the bootstrap when they are applied in a manner that requires analogous computational burden.

\section{Imputation with resampling procedures} \label{theory}

Let $\bm{\chi}$ denote the full dataset, and let $\bm{\chi}_{\rm obs}$ and $\bm{\chi}_{\rm mis}$ indicate the observed and missing portions of $\bm{\chi}$, respectively.  Imputed datasets are created by sampling $\bm{\chi}_{{\rm mis},k}$ from $P(\bm{\chi}_{\rm mis}|\bm{\chi}_{\rm obs})$ independently for $k = 1,\ldots,m$ and then setting the $k^{\rm th}$ imputed dataset as $\bm{\chi}_{k} = \{\bm{\chi}_{\rm obs};\bm{\chi}_{{\rm mis}, k}\}$.
For a  parameter of interest $\theta$, let $\widehat{\theta}_k$ denote a version of $\theta$ that is estimated using the $k^{\rm th}$ dataset, and let $\hat\nu_k$ denote an estimator of the variance of $\widehat{\theta}_k$.
\cite{rubin87a} proposes rules for combining the $\widehat\theta_k$ and $\hat\nu_k$ across $k=1,\ldots,m$ to develop interval estimates for $\theta$ that account for imputation error.

Several authors have illustrated that the Rubin combining rules do not apply
when the collected data are subject to a complex sample design \cite[e.g.,][]{kott95, kim06}. The use of a jackknife or bootstrap has been proposed as an alternative to the traditional use of multiple imputation; however, the application of such procedures in conjunction with multiple imputation has not yet been fully investigated.

\subsection{Jackknife for multiple imputation} \label{jack}

The use of a jackknife for variance estimation dates to \cite{quenouille49, quenouille56}.  Here, we examine a delete-$d$ (or delete-a-group) jackknife \citep{shao89, kott01}.
To begin, the dataset $\bm{\chi}$ is segmented into $G$ mutually exclusive and exhaustive groups denoted $\bm{\chi}^{(g)}$ for $g = 1,\ldots,G$, so that $\bm{\chi} = \{(\bm{\chi}^{(1)})';\ldots;(\bm{\chi}^{(G)})'\}'$. The replicate datasets are then notated $\bm{\chi}^{(-g)}$, which is equivalent to $\bm{\chi}$ with cases in $\bm{\chi}^{(g)}$ removed.

The applicability of a jackknife with imputed data has been considered previously \citep{rao92, righi14}; these authors recommend single imputation within each replicate group. However, they address the limited circumstance of hot deck imputation across a single imputed variable. To extend the applicability of the jackknife to more general settings, we consider multiply imputing within each replicate group; as such, we propose using multiple imputation independently within each replicate group.  Let $\hat\theta^{(-g)}$ denote the esimtate of the parameter of interest $\theta$ as found using $m$ multiply imputed versions of $\bm{\chi}^{(-g)}$.

To elaborate, for each $g = 1,\ldots, G$, the dataset $\bm{\chi}^{(-g)}$ is independently imputed $m$ times.  We let $\hat\theta^{(-g)}_j$ indicate the value of a parameter of interest $\theta$ estimated from the $j^{\rm th}$ imputed version of $\bm{\chi}^{(-g)}$.  As such, our estimate of $\theta$ for the $g^{\rm th}$ replicate group is
$ 
\hat\theta^{(-g)} = m^{-1} \sum^{m}_{j=1} \hat\theta^{(-g)}_j,
$ 
and the jackknife estimate of $\theta$ is
\begin{equation} \label{meanG}
\bar\theta_{\rm jack} = \frac{1}{G} \sum^{G}_{g=1} \hat\theta^{(-g)},
\end{equation}
the variance of which is approximated using
\begin{equation} \label{varG}
\widehat{\mbox{Var}}(\bar\theta_{\rm jack}) = \frac{G-1}{G}\sum^{G}_{g=1}(\hat\theta^{(-g)}-\bar\theta_{\rm jack})^2.
\end{equation}

To establish the validity of (\ref{meanG}) and (\ref{varG}) for inferences regarding $\theta$, we first consider the analogues of these terms for infinite $m$.  In particular, define $\theta^{(-g)}$ so that $\hat\theta^{(-g)} \overset{p}{\to} \theta^{(-g)}$ as $m \rightarrow \infty$, where $\overset{p}{\to}$ denotes convergence in probability. Then, let
\[
\bar\theta = \frac{1}{G}\sum^{G}_{g=1}\theta^{(-g)}
~~~~~\mbox{and}~~~~~
\widetilde{\mbox{Var}}(\bar\theta) = \frac{G-1}{G}\sum^{G}_{g=1}(\theta^{(-g)}-\bar\theta)^2,
\]
The terms $\bar\theta$ and $\widetilde{\mbox{Var}}(\bar\theta)$ will obey standard statistical theory for a jackknife, wherein the pseudovalues, defined as $\theta^{(g)} = G\bar\theta - (G-1)\theta^{(-g)}$, will observe $\mbox{Var}(\theta^{(g)}) \approx \omega^2$ for some constant $\omega$ so that $\mbox{Var}(\bar\theta) \approx \omega^2/G$ (and as such, $\widetilde{\mbox{Var}}(\bar\theta)$ is calculated as the sample variance of the $\theta^{(g)}$ divided by $G$).

However, for finite $m$, the imputations induce noise into the above estimators. That is, $\hat\theta^{(-g)} = \theta^{(-g)} + \epsilon^{(-g)}$, where $E[\epsilon^{(-g)}]=0$ and $\mbox{Var}(\epsilon^{(-g)}) = Gc^2/[(G-1)m]$ where $c^2>0$ is a constant giving the variance added by a single imputed version of the full sample. It follows that $\bar\theta_{\rm jack} = \bar\theta + \bar\epsilon$, where $\bar\epsilon = G^{-1} \sum^{G}_{g=1} \epsilon^{(-g)}$. Furthermore,
\[
\widehat{\mbox{Var}}(\bar\theta_{\rm jack}) = \widetilde{\mbox{Var}}(\bar\theta) + \frac{2(G-1)}{G}\sum^{G}_{g=1}(\theta^{(-g)}-\bar\theta)(\epsilon^{(-g)}-\bar\epsilon)
+ \frac{G-1}{G}\sum^{G}_{g=1}(\epsilon^{(-g)}-\bar\epsilon)^2.
\]
The second term on the right side of the above expression is negligible due to the independence of the $\epsilon^{(-g)}$ and the $\theta^{(-g)}$. Since the sample variance of the $\epsilon^{(-g)}$ is approximately equal to $Gc^2/[m(G-1)]$, we see that
\[
\widehat{\mbox{Var}}(\bar\theta_{\rm jack}) \approx \widetilde{\mbox{Var}}(\bar\theta) + \frac{G-1}{m}c^2.
\]
As such, $m$ needs to be large in comparison to $G$ in order to render the difference between $\widehat{\mbox{Var}}(\bar\theta_{\rm jack})$ and $\widetilde{\mbox{Var}}(\bar\theta)$ to be negligible. That is, $\widehat{\mbox{Var}}(\bar\theta_{\rm jack})$ diverges from $\widetilde{\mbox{Var}}(\bar\theta)$ with increasing $G$ for fixed $m$. 
(Note that this divergence is a consequence of the fact that pseudovalues defined using the $\epsilon^{(-g)}$ -- unlike the $\theta^{(-g)}$ -- have a variance that increases with $G$.)
Consequentially, the jackknife procedure will fail if single imputation \citep[$m=1$, as recommended by][p.~83]{little02} is used with probabilistic sampling of imputations.

\subsection{Bootstrap for multiple imputation} \label{boot}

An alternative to the jackknife is the bootstrap \citep{efron94}. Let $S^{(b)}$ denote a random sample, taken with replacement, of rows from $\bm{\chi}$. The cardinality of $S^{(b)}$ equals the total number of rows of $\bm{\chi}$.  Let $\bm{\chi}^{(b)}$ denote a version of $\bm{\chi}$ containing only cases in $S^{(b)}$ (with repeated rows as needed), and assume that $B$ bootstrap samples are independently created.

Akin to our approach with the jackknife, and in accordance with the observations of \cite{schomaker18}, we recommend multiply imputing each of the bootstrap samples $m$ times in order to adequately approximate each $\theta^{(b)}$.  Treating repeated rows as unique observations, let $\hat\theta^{(b)}_j$ denote the estimate of $\theta$ collected from the $j^{\rm th}$ imputed dataset for the $b^{\rm th}$ bootstrapped sample, and therefore $\hat\theta^{(b)}=m^{-1}\sum^{m}_{j=1}\theta^{(b)}_j$ indicates the best estimator of $\theta^{(b)}$.  The bootstrapped estimate of $\theta$ is then
\begin{equation} \label{eB}
\hat\theta_{\rm boot}=\frac{1}{B}\sum^{B}_{b=1}\hat\theta^{(b)},
\end{equation}
and the variance of $\hat\theta_{\rm boot}$ is approximated with
\begin{equation} \label{varB}
\widehat{\mbox{Var}}(\hat\theta_{\rm boot}) = \frac{1}{B-1} \sum^{B}_{b=1} (\hat\theta^{(b)} - \hat\theta_{\rm boot})^2.
\end{equation}

As with our discussion involving the jackknife, we consider analogues of (\ref{eB}) and (\ref{varB}) for infinite $m$. That is, define $\theta^{(b)}$ so that $\hat\theta^{(b)} \overset{p}{\to} \theta^{(b)}$ as $m \rightarrow \infty$. Let
$
\bar\theta_{\rm boot} = B^{-1}\sum^{B}_{b=1}\theta^{(b)}.
$ 
If the values of $\theta^{(b)}$ were known, the variance of $\bar\theta_{\rm boot}$ would be approximated with
\[
\widetilde{\mbox{Var}}(\bar\theta_{\rm boot}) = \frac{1}{B-1} \sum^{B}_{b=1} (\theta^{(b)} - \bar\theta_{\rm boot})^2.
\]
The terms $\bar\theta_{\rm boot}$ and $\widetilde{\mbox{Var}}(\bar\theta_{\rm boot})$ will obey standard theory for a bootstrap as valid estimators for inference of $\theta$, and as such, the discrepancy between $\widehat{\mbox{Var}}(\hat\theta_{\rm boot})$ and $\widetilde{\mbox{Var}}(\bar\theta_{\rm boot})$ is of interest.

Note that $\hat\theta^{(b)} = \theta^{(b)} + \epsilon^{(b)}$ where $E[\epsilon^{(b)}]=0$ and $\mbox{Var}(\epsilon^{(b)})=c^2/m$, where $c^2$ was defined in Section \ref{jack}.  Letting $\bar\epsilon_{\rm boot} = B^{-1}\sum^{B}_{b=1}\epsilon^{(b)}$, it follows that
\[
\widehat{\mbox{Var}}(\bar\theta_{\rm boot}) = \widetilde{\mbox{Var}}(\bar\theta_{\rm boot}) + \frac{2}{B-1}\sum^{B}_{b=1}(\theta^{(b)}-\bar\theta_{\rm boot})(\epsilon^{(b)}-\bar\epsilon_{\rm boot})
+ \frac{1}{B-1}\sum^{B}_{b=1}(\epsilon^{(b)}-\bar\epsilon_{\rm boot})^2.
\]
The second term in the above is negligible due to the independence of the $\theta^{(b)}$ and $\epsilon^{(b)}$, whereas the third term is simply the sample variance of the $\epsilon^{(b)}$. Consequentially,
\[
\widehat{\mbox{Var}}(\bar\theta_{\rm boot}) \approx \widetilde{\mbox{Var}}(\bar\theta_{\rm boot}) + \frac{c^2}{m},
\]
and as such the discrepancy between $\widehat{\mbox{Var}}(\hat\theta_{\rm boot})$ and $\widetilde{\mbox{Var}}(\bar\theta_{\rm boot})$ decreases with increasing $m$ and in effect does not depend upon $B$.

The bootstrap is believed to require more bootstrapped samples ($B$) than the jackknife requires replication groups ($G$). For instance, a rule-of-thumb of $B=40n$ has been proposed in settings involving fully observed data \citep{davison97}. However, since (unlike the jackknife) $\hat\theta^{(b)}$ is not divergent from $\theta^{(b)}$ with increasing $B$ for fixed $m$, the bootstrap should not require $m$ to be as large as is needed with the jackknife.

\section{Simulations} \label{sims}

We next present a brief simulation study to verify the theory developed in Section \ref{theory}. For this purpose, we consider a setup where resampling procedures are the preferred method of variance estimation in survey data that may be subject to missingness.  Specifically, we consider the use of propensity score-type inverse probability weighting for the purpose of blending probability and convenience samples as described by \cite{robbins21}. Therein, it is shown that traditional linearization techniques fail in such settings, mandating the use of replication procedures for variance estimation.

For $i \in \Omega$ where $\Omega$ denotes a synthetic population, we let $Y_i$ (the only variable subject to missingness) denote an outcome of interest, $X_{1i}$ denote an auxiliary variable that governs the missingness mechanism in $Y_i$, and $X_{2i}$ denote an auxiliary variable that governs discrepancies between the probability and convenience samples.

Continuing, $X_{1i}$ and $X_{2i}$ are sampled from a bivariate normal distribution with zero means, unit variances, and a correlation of 0.5, and $Y_i$ is generated as a binary random variable with success probabilty $p_{y,i}=1/(1+\exp\{-X_{1i}+X_{2i}\})$. A population of size $N=20,000$ is independently generated according to these constructs. Next, each unit in the population is assigned to the probability sample with a probability $p_{s,i}=0.02$, and each unit not assigned to the probability sample is assigned to the convenience sample with probability $p_{c,i}=1/(1+\exp\{10/3-0.75X_{2i}\})$. Lastly, each value of $Y_i$ is missing with probability $p_{m,i}=1/(1+\exp\{-2/3-0.5X_{1i}\})$ (for a missingness rate of approximately 33\%).

Following the simultaneous propensity score blending methodology proposed by \cite{robbins21}, we let $\hat\gamma_i$ denote the estimated probability that case $i$ is in the convenience sample, conditional on $X_{1i}$ and $X_{2i}$ and given that the case is in either the probability or convenience samples (this probability is estimated here using a logistic regression model). The weight for case $i$ is set to $w_i = (1-\hat\gamma_i)/p_{s,i}$.  We aim to estimate $\mu = E[Y_i]$, which would be performed using $\hat\mu = \sum_{i \in S}w_iY_i/\sum_{i \in S}w_i$ if $Y_i$ were observed for all $i \in S$, where $S$ denotes the union of the probability and convenience samples.

Let $S^{(r)}$ denote a replicate version of $S$ for a jackknife or bootstrap procedure (i.e., $S^{(-g)}$ or $S^{(b)}$).  Missing values of $Y_i$ for $i \in S^{(r)}$ are imputed $m$ times, and let $Y^{(r)}_{i,k}$ denote the $k^{\rm th}$ imputed value of $Y_I$ for $i \in S^{(r)}$. Furthermore, in line with the recommendations of \cite{robbins21} and others, weights are recalculated for each replicate group, yielding $w_{i}^{(r)}$ for $i \in S^{(r)}$. In this case, $\hat\mu^{(r)}$, the multiple imputation estimate of $\mu$ for $S^{(r)}$, is calculated via
\[
\hat\mu^{(r)} =\frac{1}{m}\sum^m_{k=1}\sum_{i \in S^{(r)}}w^{(r)}_iY^{(r)}_{i,k}/\sum_{i \in S^{(r)}}w^{(r)}_i.
\]
Jackknife and bootstrap estimates of $\mu$, along with their approximate variance, are then calculated in line with (\ref{meanG})-(\ref{varB}). Imputations for missing $Y_i$ are created independently $m$ times for each $g=1,\ldots,G$ and $b=1,\ldots,B$ using 
\texttt{mice} \citep{vanbuuren10} with imputation by logistic regression and with 5 iterations of Markov chain Monte Carlo. In accordance with common practice for imputing with weighted data \citep{schenker06, robbins23b}, the imputation model includes $X_1$ and $X_2$ and the weights.

The above data generation process is repeated $J = 2000$ times. For a given method, we let $\tilde\mu^{[j]}$ denote the point estimate of $\mu$ estimated at the $j^{\rm th}$ simulation (e.g., $\tilde\mu^{[j]}$ for a bootstrap is the average of the respective $\hat\mu^{(b)}$ for $b=1,\ldots,B$).
For each method, we calculate the bias and root mean-squared error (rMSE) in the estimate of $\mu$ as follows:
\[
\mbox{bias}_\mu = \frac{1}{J}\sum^{J}_{j=1}(\tilde\mu^{[j]}-\mu),
~~~~~~\mbox{and}~~~~
\mbox{rMSE}_\mu = \sqrt{\frac{1}{J}\sum^{J}_{j=1}(\tilde\mu^{[j]}-\mu)^2}.
\]
We denote the corresponding estimate for $\mbox{Var}(\tilde\mu^{[j]})$ with $\tilde{V}^{[j]}$. The $100(1-\alpha)^{\rm th}$ confidence interval for $\mu$ is then calculated with bounds given by
$ 
\tilde\mu^{[j]} \pm z_{1-\alpha/2} \sqrt{\tilde{V}^{[j]}},
$ 
where $z_{1-\alpha/2}$ is the $100(1-\alpha)^{\rm th}$ percentile of the standard normal distribution. 
The coverage for each method is calculated as
the portion of simulations in which the confidence interval contains the true value of $\mu$.

For the purposes of comparison, we also consider a simplified method where $m$ imputed versions of each synthetic dataset are generated, and then the respective jackknife and bootstrap estimates are created using the original imputations for that dataset without re-imputing at each replicate group. Although this is not appropriate in theory (since information from the full dataset factors into each replication estimate), it will be substantially less computationally burdensome than the full methods proposed above.
Results for the for the simulation process outlined above are shown in Table \ref{tab1}.

\begin{table}[!ht]
\caption{\label{tab1}Simulated bias, rMSE, and coverage (using $1-\alpha=0.95$) for various $m$ and $G$ or $B$ for the mean of $Y_i$ when imputations are resampled for each replication group and when a single batch of $m$ imputated datasets is used for across all replication groups.
\vspace{.08in}
}
\centering
\setlength{\tabcolsep}{.4em}
\begin{tabular}{clccccccc}
\hline \hline
& & \multicolumn{3}{c}{\specialcell{Re-impute for each \\ replication group}} & & \multicolumn{3}{c}{\specialcell{Use same imputations \\ across replication groups}} \\ \cline{3-5} \cline{7-9}
& & Bias & rMSE & Cover. & & Bias & rMSE & Cover. \\  \hline
\multirow{8}{*}{\specialcell{Jackknife \\ ($G=25$)}} & $m = 1$ & 0.0002 & 0.0216 & 1.0000 &  & 0.0001 & 0.0238 & 0.7875 \\ 
& $m = 5$ & 0.0002 & 0.0215 & 0.9870 &  & 0.0002 & 0.0220 & 0.7630 \\ 
& $m = 10$ & 0.0002 & 0.0215 & 0.9680 &  & 0.0002 & 0.0217 & 0.7640 \\ 
& $m = 25$ & 0.0001 & 0.0215 & 0.9555 &  & 0.0001 & 0.0217 & 0.7620 \\ 
& $m = 50$ & 0.0001 & 0.0215 & 0.9500 &  & 0.0001 & 0.0216 & 0.7575 \\ 
& $m = 100$ & 0.0001 & 0.0215 & 0.9410 &  & 0.0001 & 0.0215 & 0.7580 \\ 
& $m = 150$ & 0.0001 & 0.0215 & 0.9420 &  & 0.0001 & 0.0215 & 0.7590 \\ 
& $m = 200$ & 0.0001 & 0.0215 & 0.9400 &  & 0.0001 & 0.0215 & 0.7630 \\  \hline 
\multirow{5}{*}{\specialcell{Bootstrap \\ ($B=25$)}} & $m = 1$ & 0.0002 & 0.0220 & 0.9515 &  & 0.0001 & 0.0239 & 0.7870 \\ 
& $m = 5$ & 0.0002 & 0.0219 & 0.9275 &  & 0.0002 & 0.0222 & 0.7620 \\ 
& $m = 10$ & 0.0002 & 0.0218 & 0.9295 &  & 0.0001 & 0.0219 & 0.7590 \\ 
& $m = 25$ & 0.0002 & 0.0219 & 0.9285 &  & 0.0001 & 0.0218 & 0.7465 \\ 
& $m = 50$ & 0.0002 & 0.0219 & 0.9285 &  & 0.0001 & 0.0217 & 0.7515 \\  \hline 
\multirow{5}{*}{\specialcell{Bootstrap \\ ($B=100$)}} & $m = 1$ & 0.0002 & 0.0216 & 0.9645 &  & 0.0001 & 0.0238 & 0.7940 \\ 
& $m = 5$ & 0.0002 & 0.0216 & 0.9465 &  & 0.0001 & 0.0220 & 0.7765 \\ 
& $m = 10$ & 0.0002 & 0.0216 & 0.9435 &  & 0.0001 & 0.0217 & 0.7750 \\ 
& $m = 25$ & 0.0002 & 0.0216 & 0.9410 &  & 0.0001 & 0.0217 & 0.7695 \\ 
& $m = 50$ & 0.0002 & 0.0216 & 0.9405 &  & 0.0000 & 0.0216 & 0.7680 \\  \hline 
\multirow{5}{*}{\specialcell{Bootstrap \\ ($B=250$)}} & $m = 1$ & 0.0001 & 0.0216 & 0.9655 &  & 0.0001 & 0.0238 & 0.7975 \\ 
& $m = 5$ & 0.0001 & 0.0216 & 0.9485 &  & 0.0001 & 0.0220 & 0.7755 \\ 
& $m = 10$ & 0.0001 & 0.0216 & 0.9480 &  & 0.0001 & 0.0218 & 0.7745 \\ 
& $m = 25$ & 0.0001 & 0.0216 & 0.9460 &  & 0.0000 & 0.0217 & 0.7725 \\ 
& $m = 50$ & 0.0001 & 0.0216 & 0.9445 &  & 0.0000 & 0.0216 & 0.7740 \\  \hline 
\end{tabular} 
\end{table}

The findings illustrate that for both the jackknife and the bootstrap, it is necessary to impute independently for each replication group and to use multiple imputations. As suggested by the theoretical results in the prior section, we see that a substantially larger $m$ is required for for the jackknife than the bootstrap. However, there are also indications that the value of $B$ should be larger for the bootstrap than the corresponding value of $G$ for the jackknife---as evidenced by the larger rMSE for the bootstrap with smaller values of $B$ are used. Note, likewise, that the jackknife with $G=25$ and $m=200$ yields a slightly smaller rMSE than the bootstrap with $B=250$ and $m=50$; this helps to justify the use of the jackknife in spite of it requiring a larger $m$. 

\section{Discussion} \label{sec5}

Both imputation and replication procedures are of fundamental importance in survey analysis. This note elucidates the steps that must be taken in order to draw correct inferences when using these procedures in tandem. We contradict some conventional wisdom \citep[e.g.,][]{little02} by illustrating that for both a jackknife and bootstrap, each replication group should be multiply imputed independently of other groups whenever stochastic sampling of imputations is used. 

In applications with fullly observed data, the bootstrap has been argued to outperform a jackknife for general purposes \citep{efron81, efron82}, although a bootstrap is believed to be more computationally intensive since the number of bootstrapped samples required is argued to be substantial \citep{zoubir07}. This observation seemingly increases the appeal of the jackknife in our setting since the need to multiply impute each replication group enhances the computational burden of both procedures substantially. However, we illustrate that the number of multiple imputations per replication groups needed increases with the number of groups for the jackknife but not the bootstrap. Our simulations suggest that when the two methods are applied in a manner that mandates a similar computational load, the jackknife may yield a slightly smaller rMSE.


\setlength{\bibsep}{0.0pt}
\singlespacing
\bibliographystyle{Chicago}

\bibliography{arxivRef}

\begin{thebibliography}{}

\bibitem[\protect\citeauthoryear{Binder}{Binder}{1983}]{binder83}
Binder, D.~A. (1983).
\newblock On the variances of asymptotically normal estimators from complex
  surveys.
\newblock {\em International Statistical Review/Revue Internationale de
  Statistique\/}~{\em 51\/}(3), pp. 279--292.

\bibitem[\protect\citeauthoryear{Davison and Hinkley}{Davison and
  Hinkley}{1997}]{davison97}
Davison, A.~C. and D.~V. Hinkley (1997).
\newblock {\em Bootstrap methods and their application}.
\newblock Number~1. Cambridge university press.

\bibitem[\protect\citeauthoryear{Efron}{Efron}{1981}]{efron81}
Efron, B. (1981).
\newblock Nonparametric estimates of standard error: the jackknife, the
  bootstrap and other methods.
\newblock {\em Biometrika\/}~{\em 68\/}(3), 589--599.

\bibitem[\protect\citeauthoryear{Efron}{Efron}{1982}]{efron82}
Efron, B. (1982).
\newblock {\em The jackknife, the bootstrap and other resampling plans}.
\newblock SIAM.

\bibitem[\protect\citeauthoryear{Efron and Tibshirani}{Efron and
  Tibshirani}{1994}]{efron94}
Efron, B. and R.~J. Tibshirani (1994).
\newblock {\em An introduction to the bootstrap}.
\newblock CRC press.

\bibitem[\protect\citeauthoryear{Kim, Michael~Brick, Fuller, and Kalton}{Kim
  et~al.}{2006}]{kim06}
Kim, J.~K., J.~Michael~Brick, W.~A. Fuller, and G.~Kalton (2006).
\newblock On the bias of the multiple-imputation variance estimator in survey
  sampling.
\newblock {\em Journal of the Royal Statistical Society Series B: Statistical
  Methodology\/}~{\em 68\/}(3), 509--521.

\bibitem[\protect\citeauthoryear{Kott}{Kott}{1995}]{kott95}
Kott, P. (1995).
\newblock A paradox of multiple imputation.
\newblock In {\em Proceedings of the Section on Survey Research Methods}, pp.\
  384--389. American Statistical Association.

\bibitem[\protect\citeauthoryear{Kott}{Kott}{2001}]{kott01}
Kott, P.~S. (2001).
\newblock The delete-a-group jackknife.
\newblock {\em Journal of Official Statistics\/}~{\em 17\/}(4), 521.

\bibitem[\protect\citeauthoryear{Little and Rubin}{Little and
  Rubin}{2002}]{little02}
Little, R.~J.~A. and D.~B. Rubin (2002).
\newblock {\em Statistical {A}nalysis with {M}issing {D}ata\/} (2nd ed.).
\newblock New {J}ersey: John {W}iley \& {S}ons.

\bibitem[\protect\citeauthoryear{Lumley}{Lumley}{2004}]{lumley04}
Lumley, T. (2004).
\newblock Analysis of complex survey samples.
\newblock {\em Journal of Statistical Software\/}~{\em 9\/}(1), 1--19.

\bibitem[\protect\citeauthoryear{Miller and Kott}{Miller and
  Kott}{2011}]{miller11}
Miller, D. and P.~Kott (2011).
\newblock Using the {DAG} jackknife to measure the variance of an estimator in
  the presence of item nonresponse.
\newblock {\em JSM Proceedings, Statistical Computing Section. Alexandria, VA:
  Am. Statist. Assoc\/}.

\bibitem[\protect\citeauthoryear{Quenouille}{Quenouille}{1949}]{quenouille49}
Quenouille, M.~H. (1949).
\newblock Problems in plane sampling.
\newblock {\em The Annals of Mathematical Statistics\/}~{\em 20\/}(3),
  355--375.

\bibitem[\protect\citeauthoryear{Quenouille}{Quenouille}{1956}]{quenouille56}
Quenouille, M.~H. (1956).
\newblock Notes on bias in estimation.
\newblock {\em Biometrika\/}~{\em 43\/}(3/4), 353--360.

\bibitem[\protect\citeauthoryear{Rao and Shao}{Rao and Shao}{1992}]{rao92}
Rao, J.~N. and J.~Shao (1992).
\newblock Jackknife variance estimation with survey data under hot deck
  imputation.
\newblock {\em Biometrika\/}~{\em 79\/}(4), 811--822.

\bibitem[\protect\citeauthoryear{Righi, Falorsi, and Fasulo}{Righi
  et~al.}{2014}]{righi14}
Righi, P., S.~Falorsi, and A.~Fasulo (2014).
\newblock A modified extended delete a group jackknife variance estimator under
  random hot deck imputation in business surveys.
\newblock In {\em Contributions to Sampling Statistics}, pp.\  219--233.
  Springer.

\bibitem[\protect\citeauthoryear{Robbins}{Robbins}{2023}]{robbins23b}
Robbins, M.~W. (2023).
\newblock {Joint Imputation of General Data}.
\newblock {\em Journal of Survey Statistics and Methodology\/}, smad034.

\bibitem[\protect\citeauthoryear{Robbins, Ghosh-Dastidar, and Ramchand}{Robbins
  et~al.}{2021}]{robbins21}
Robbins, M.~W., B.~Ghosh-Dastidar, and R.~Ramchand (2021).
\newblock Blending probability and nonprobability samples with applications to
  a survey of military caregivers.
\newblock {\em Journal of Survey Statistics and Methodology\/}~{\em 9\/}(5),
  1114--1145.

\bibitem[\protect\citeauthoryear{Rubin}{Rubin}{1987}]{rubin87a}
Rubin, D.~B. (1987).
\newblock {\em {Multiple Imputation for Nonresponse in Surveys}}.
\newblock New {Y}ork, {N}ew {Y}ork: John {W}iley \& {S}ons.

\bibitem[\protect\citeauthoryear{Rubin}{Rubin}{1996}]{rubin96}
Rubin, D.~B. (1996).
\newblock Multiple imputation after 18+ years.
\newblock {\em Journal of the {A}merican {S}tatistical {A}ssociation\/}~{\em
  91}, 473--489.

\bibitem[\protect\citeauthoryear{Schenker, Raghunathan, Chiu, Makuc, Zhang, and
  Cohen}{Schenker et~al.}{2006}]{schenker06}
Schenker, N., T.~E. Raghunathan, P.-L. Chiu, D.~M. Makuc, G.~Zhang, and A.~J.
  Cohen (2006).
\newblock Multiple imputation of missing income data in the {N}ational {H}ealth
  {I}nterview {S}urvey.
\newblock {\em Journal of the {A}merican {S}tatistical {A}ssociation\/}~{\em
  101}, 924--933.

\bibitem[\protect\citeauthoryear{Schomaker and Heumann}{Schomaker and
  Heumann}{2018}]{schomaker18}
Schomaker, M. and C.~Heumann (2018).
\newblock Bootstrap inference when using multiple imputation.
\newblock {\em Statistics in Medicine\/}~{\em 37\/}(14), 2252--2266.

\bibitem[\protect\citeauthoryear{Shao and Wu}{Shao and Wu}{1989}]{shao89}
Shao, J. and C.~J. Wu (1989).
\newblock A general theory for jackknife variance estimation.
\newblock {\em The Annals of Statistics\/}, 1176--1197.

\bibitem[\protect\citeauthoryear{Van~Buuren and Groothuis-Oudshoorn}{Van~Buuren
  and Groothuis-Oudshoorn}{2010}]{vanbuuren10}
Van~Buuren, S. and K.~Groothuis-Oudshoorn (2010).
\newblock mice: Multivariate imputation by chained equations in r.
\newblock {\em Journal of Statistical Software\/}, 1--68.

\bibitem[\protect\citeauthoryear{Zoubir and Iskandler}{Zoubir and
  Iskandler}{2007}]{zoubir07}
Zoubir, A.~M. and D.~R. Iskandler (2007).
\newblock Bootstrap methods and applications.
\newblock {\em IEEE Signal Processing Magazine\/}~{\em 24\/}(4), 10--19.

\end{thebibliography}

\end{document}